\documentclass[a4paper]{spie}  
\usepackage[]{graphicx}

\usepackage[left=1.925cm, right=1.925cm, top=2.54cm, bottom=4.94cm]{geometry}

\title{Imaging reconstruction for infrared interferometry: first images of YSOs environment} 

\author{S. Renard\supit{a}, F. Malbet\supit{a}, E. Thi\'ebaut\supit{b},
  J.-P. Berger\supit{a}
\skiplinehalf
\supit{a}Laboratoire d'Astrophysique de Grenoble
(LAOG), Universit\'e Joseph Fourier/CNRS, BP 53, F-38041 Grenoble, France; \\
\supit{b}Centre de Recherche Astronomique de Lyon (CRAL), Observatoire
de Lyon,  Universit\'e
Claude Bernard, France.
}

\authorinfo{Further author information: please send correspondence to
  S.R. \texttt{Stephanie.Renard@obs.ujf-grenoble.fr}}

\DeclareTextSymbol{\degre}{OT1}{23}

\begin{document} 
  \maketitle 

\begin{abstract}

  The study of protoplanetary disks, where the planets are believed to
  form, will certainly allow the formation of our Solar System to be
  understood. To conduct observations of these objects at the
  milli-arcsecond scale, infrared interferometry provides the right
  performances for T Tauri, FU Ori or Herbig Ae/Be stars.  However,
  the only information obtained so far are scarce visibility
  measurements which are directly tested with models. With the outcome
  of recent interferometers, one can foresee obtaining images
  reconstructed independently of the models. In fact, several
  interferometers including IOTA and AMBER on the VLTI already provide the
  possibility to recombine three telescopes at once and thus to obtain
  the data necessary to reconstruct images.

  In this paper, we describe the use of MIRA, an image reconstruction
  algorithm developed for optical interferometry data (squared
  visibilities and closure phases) by E.~Thi\'ebaut. We foresee also
  to use the spectral information given by AMBER data to constrain
  even better the reconstructed images. We describe the use of
  MIRA to reconstruct images of young stellar objects out of actual
  data, in particular the multiple system GW Orionis (IOTA, 2004), and
  discuss the encountered difficulties.

\end{abstract}


\section{INTRODUCTION}
\label{sec:intro}

The process of star formation triggered by the collapse and
fragmentation of a molecular cloud is believed to lead to a protostar
surrounded by an accreting disk undergoing important mass lost
(wind, jets).  The study of the circumstellar disks during this
pre-main sequence evolution is important because these disks are
probably the birth location of future planetary systems. Therefore,
the understanding of their physical conditions is necessary before we
can understand the planets formation process.  The details of the
physical processes which are taking place in the circumstellar disk
are not well known because of a lack of data to constrain them.  The
spatial distances at which the planetary, gas accretion and wind/jet
formation occur, correspond to astronomical unit scales requiring
milli-arcesecond spatial resolution to be probed unambiguously.
Therefore, we used infrared interferometry which is one of the
techniques capable to reach this spatial resolution at those
wavelengths.

Interferometry consists of a coherent beam combination of the light
received from an astronomical source and collected by two or more
independent telescopes. As the light beams overlapped, we observe an
interference pattern which is composed of fringes. An ideal
interferometer measures the complex visibility $V(u,v)$ which is,
according to the Van Cittert-Zernicke theorem, the Fourier transform
of the object brightness distribution $I(x,y)$.  In reality, the
atmosphere turbulence induces random phase shifts and this phenomena
does allow the Fourier phases to be properly calibrated. Therefore, we
need to compute interferometric measurements that are insensitive to
this perturbation :
\begin{itemize}
 \item the squared visibility : $V^{(2)}(u,v) = \hspace{0.2cm} \mid~V(u,v)~\mid^2$
 \item the bispectrum : $B(u_i,u_j,v_i,v_j) =
   V(u_i,v_i) \times V(u_j,v_j) \times V(-u_i-u_j,-v_i-vj)$. The phase of the
   bispectrum is called the closure phase (CP), independant of the
   atmosphere delay phase. This method allows us to retrieve a part of the
   Fourier phase.
\end{itemize}

From these interferometric data, the astronomers up to now are fitting
simple parametric or homemade models in order to directly constrain
their parameters to these visibilities and closure phases and in order
to better understand the structure of the inner part of the disk.
Yet, although important observations from these analyzes have lead to
a profound revision of the inner disk structure at spatial scale
corresponding to the planet formation, the lack of model independent
imaging, without any knowledge of the observed structure, still
forbids unambiguous diagnostics of the physical processes at work.

With the outcome of interferometers able to combine 3 beams, like
IONIC on IOTA and AMBER on the VLTI, and later of full imaging array
like the second generation of instruments at VLTI, one foresees
to obtain images reconstructed independently of the models. In fact,
the AMBER instrument at the VLTI provides the possibility to recombine
three telescopes at once and thus to obtain many data necessary to
reconstruct images.

\section{METHOD} 

In this section we present the method for performing image
reconstruction. More details is given by Thi\'ebaut\cite{ThiebautSPIE}
in this volume.

\subsection{Image reconstruction in interferometry} 
\label{sec:img_rec}

The observed bright distribution in a direction $s'$ of an incoherent source is given by : 
\begin{equation}
 I_{\rm obs}(s') = PSF(x,y)*I_{\rm true}(x,y)+N(x,y)
\label{eq:image_form}
\end{equation}
where $PSF(x,y)$ is the point spread function, $I_{\rm true}(x,y)$ is the true object brightness distribution and $N(x,y)$ is the noise; ``$*$'' denotes convolution. As interferometry does not measure in the image plane but in the Fourier space, we can compute the Fourier transform of the Eq.~(\ref{eq:image_form}): 
\begin{equation}
 V_{\rm obs}(u,v) = V_{\rm true}(u,v) \times S(u,v) + N'(u,v)
\label{eq:image_form_TF} 
\end{equation}
where $S(u,v)$ is the Fourier transform of the point spread function,
called the sampling function, $V_{\rm true}(u,v)$ is the Fourier
transform of the true object brightness of distribution, called the
true visibility and $N'(u,v)$ is the noise in the Fourier space.

The goal of the image reconstruction is allow to retrieve an image
that can be trusted to be close to the real emission ($I_{\rm approx}
\approx I_{\rm true}$). To achieve this goal, the method is based on a
Bayesian approach: to find a solution $z$ which has the
maximum probability given the data. Such solution is called the
maximum ``a posteriori'' (MAP)\cite{Thiébaut2006}:
\begin{equation}
 z^{\rm (MAP)} = \arg \min_{z} f(z)
\end{equation}
where the penalty function reads : 
\begin{equation}
 f(z) = f_{\rm data}(z) + \mu f_{\rm prior}(z)
\label{eq:penalty}
\end{equation}
This equation shows that we had to minimize a sum of two functions : 
\begin{itemize}
 \item the likelihood penalty term $f_{\rm data}(z)$ which measures the compatibility of the solution $z$ with the data. Typically, the data penalties are defined assuming a Gaussian statistics and it leads to a classical computation of a $\chi^2$:
\begin{equation}
  \chi^2_{V^2} = \sum \left\{ \frac {\vert
    V^2_{\rm true } - V^2_{\rm obs}\vert^2}{\sigma_{V^2}^2} \right\} 
\end{equation}
\begin{equation}
  \chi^2_{\mathrm{T}} = \sum \frac{\left\vert
    \exp(i\,\phi^{\mathrm{data}}_\mathrm{T}) -
    \exp(i\,(\varphi_1 + \varphi_2 - \varphi_3))
  \right\vert^2}{\sigma^2_{\mathrm{T}}}
\end{equation}
 \item the prior penalty term $f_{\rm prior}(z)$ which allows us to account additional constraints not carried out by the data alone. In fact, there is more unknown parameters than measurements and an infinity of solutions are in concordance with the data. The regularization allows us to choice among all the compatible solutions the one which is the closest to an ``a priori''. Because the noise usually contaminates the high frequencies, the smoothness prior is one of the most common regularization constraint: 
\begin{equation}
f_{\mathrm{prior}}(x;w) = \sum_k w_k \, \vert \hat{x}_k \vert^2
\label{eq:prior}
\end{equation}
with $\hat{x}_k$ the discrete Fourier transform of $x$ at $k$-th discrete spatial frequency and where the weights $w$ are chosen to ensure spectral smoothing to enforce smoothness of the Fourier Spectrum and thus compactness of the brightness distribution of the field of view.
\end{itemize}

To control the importance of the priory penalty according to the likelihood penalty, a weight factor $\mu$ (see Eq.~ \ref{eq:penalty}) is added in the equation. In practise, the best solution leads to $ f_{data}(z) = \eta $ where $\eta$ is the number of data. If $f_{data}(z)$ is small than $\eta$, we should increase $\mu$ and if $f_{data}(z)$ is larger than $\eta$, we should decrease $\mu$.

\subsection{Multi-Aperture Image Reconstruction algorithm (MIRA)} 
\label{sec:mira}

The Multi-Aperture Image Reconstruction Algorithm (MIRA) is an image
reconstruction algorithm devoted to optical interferometry data
developed by Thi\'ebaut\cite{ThiebautSPIE}.  Because of the small
number of telescopes, optical interferometers are only able to measure
very few Fourier frequencies at the same time and even less Fourier
phases because of the atmosphere turbulence. The solution adopted in
the MIRA algorithm is to directly fit the interferometric observables
(the squared visibilities and the closure phases) without explicity
rebuilding the missing Fourier phases, unlike
WISARD\cite{2005OptL...30.1809M}. The image reconstruction in MIRA is
regularized by using various kind of regularization (entropy,
quadratic and non-quadratic smoothness ...).  MIRA states the image reconstruction as an inverse
problem solved by minimizing a so-called penalty function (see
Sect.~\ref{sec:img_rec}).  Owing to the specific relationship between
the object brightness distribution and the interferometric data, the
penalty to minimize cannot be guaranteed to be convex. At this time,
the algorithm for the optimisation used in MIRA is the VMLM-B
non-linear constrained optimization algorithm\cite{Thiebaut02} and which imposes positivity and normalization of the sought image but only yields a local minimum. Hence the initial image of the iterative algorithm is a deciding parameter of the algorithm.

\section{IMAGE RECONSTRUCTION ON YOUNG STELLAR OBJECTS} 

In this section, we present the results obtained on two objects.  The
first one is a known young binary, GW~Orionis, and the other one is a
young star which is believed to have disk, HD~45677.

\subsection{GW Orionis} 
\label{sec:gwori}

\subsubsection{Characteristics of GW Ori}

\begin{table}[tp]
 \caption{Characteristics of the classical T Tauri star GW Orionis}
  \label{tab:gwori_caract}
   \begin{center}
    \begin{tabular}{ccccccccc}
    \hline
    \hline
     RA (J2000) & Dec (J2000) & m$_V$ & m$_J$ & m$_H$ & m$_K$ & Dist. (pc) & Spec. type \\
     \hline
     5h39m08s & +11\degre52'12.7'' & 9.9 & 7.7 & 7.1 & 6.6 & 300 & K3V  \\
     \hline
    \end{tabular}
   \end{center}
\end{table}

GW Orionis (see characteristics on Table~\ref{tab:gwori_caract}) is a
young stellar object and belongs to the classical T Tauri group.
Mathieu et al.\cite{Mathieu91} discovered that this star was in fact a
spectroscopic binary with an orbital period of 242 days. The masses of
the binary component is 2.5 solarmasses for the primary and between
0.5 and 1 solarmasses for the secondary and the separation is slightly
more than 1 AU. The center-of-mass velocity has been observed to vary
over a period of 1000 days, suggesting the presence of a third star in
the system or an $m=1$ perturbation in the associated disk.

The observed spectral energy distribution (SED) of GW Orionis shows a
large near- and far-infrared excess over the stellar photosphere and
presents two characteristics :
\begin{enumerate}
\item the morphology of the SED is double peaked with a minimum in the
  continuum near 10 $\mu$m;
\item there is a very strong 10 $\mu$m silicate emission feature.
\end{enumerate}

The interferometric data on GW Orionis used for the image
reconstruction (see Fig.~\ref{fig:gwori_graph} in red squares for the data and Fig.~\ref{fig:gwori_results} on the left panel for the $(u,v)$ plane) have
been taken at the IOTA interferometer\cite{Berger05}.  Located on
Mount Hopkins in Arizona, IOTA is a long baseline interferometer that
observes at visible and NIR wavelengths. It has three telescopes
movable among 17 stations along two orthogonal linear arms. By
observing a target in many different array configurations, IOTA can
synthesize an aperture of $35 \times 15$ m (corresponding to an
angular resolution of $5 \times 12$ mas at 1.65 $\mu$m).

\subsubsection{Reconstructed image with MIRA}

The image reconstruction of GW Ori has been done with the following
parameters:
\begin{itemize}
 \item a pixel size of 0.1 mas to have a more precise image
 \item an image width of 400 pixels to have a sufficient field of view
 \item a regularization of type ``smoothness''
\end{itemize}

\begin{figure}[tp]
  \begin{center}
    \begin{tabular}{cc}
      \includegraphics[width=0.4\textwidth]{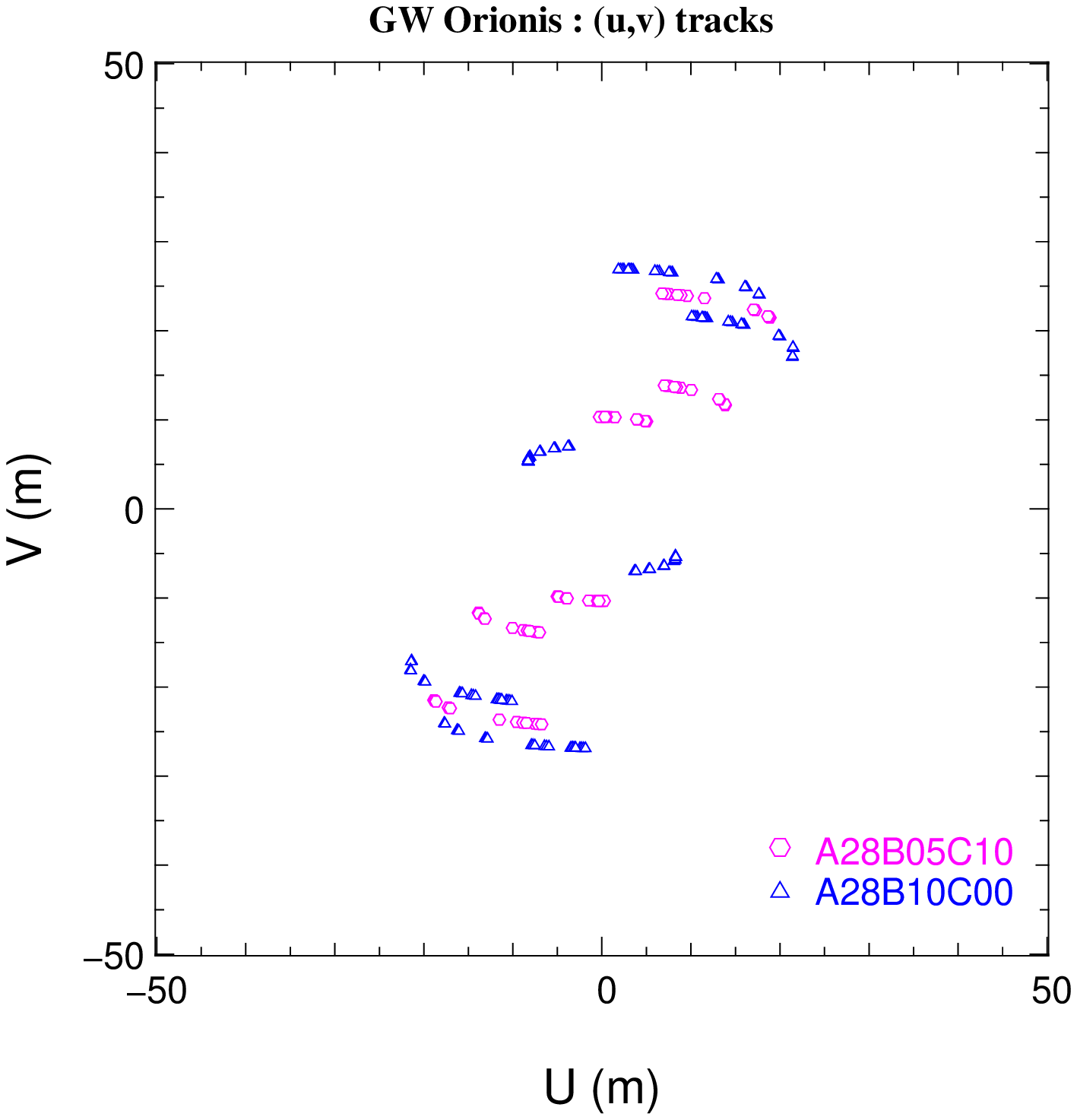} &
      \includegraphics[width=0.5\textwidth]{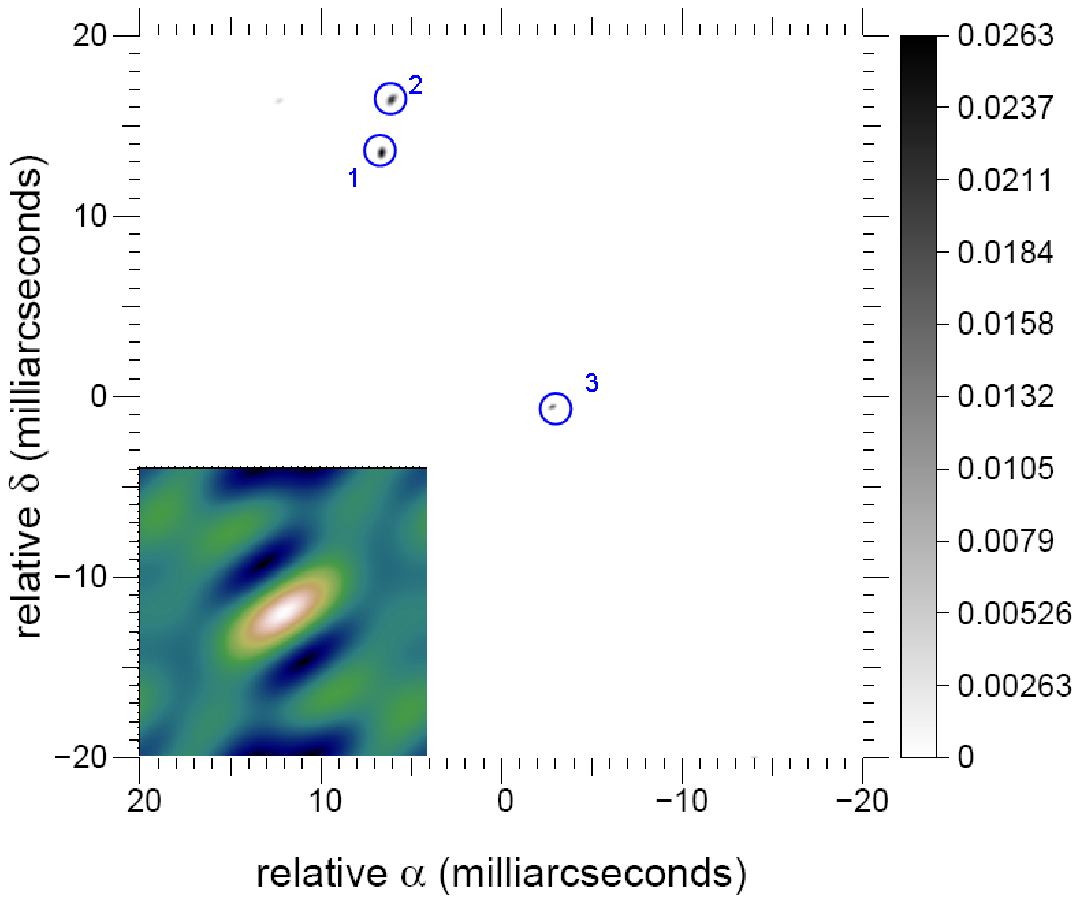}
    \end{tabular}
  \end{center}
  \caption{\label{fig:gwori_results} $(u,v)$ coverage of the data set
    and the result of the reconstructed image by MIRA for GW Ori (with the dirty map). 
    }
\end{figure}

The result of the image reconstruction is shown in
Fig.~\ref{fig:gwori_results}. Besides the image, MIRA gives also a
comparison between the measured interferometric data (squared
visibilities and closure phases) and the reconstructed interferometric
data computed by the Fourier transformation of the image, which can
give us an approximation of the $\chi^2$ and show which data point are
not matched by the image reconstruction (see Fig.~\ref{fig:gwori_graph} in green).

We find that the GW Ori system appears to be a triple system, with a
binary already known (see section above) and a faint third component.
From this image, we can measure the different parameters of the triple
system, namely the flux ratio between a reference (taken as the
brightest one) and the two other components, the distance between the
components and the position angle. The parameters measured on the
image are shown in line 1 of Table~\ref{tab:gwori_result}.

\begin{table}[tp]
 \caption{Results of the different methods used on the GW Ori data, 1 indicates the reference component, 1-2 the thight binary and 1-3 the wide binary.}
  \label{tab:gwori_result}
   \begin{center}
    \begin{tabular}{p{1.5cm}|ccc|ccc|c}
    \hline
    \hline
     & Flux ratio & dist. (mas) & pos. angle (\degre) &  Flux ratio & dist. (mas) & pos. angle (\degre) &$\chi^2$\\
     & 1-2 & 1-2 & 1-2 & 1-3 & 1-3 & 1-3 &norm.\\
     \hline
     MIRA & 0.81 & 2.97 & -10 &  0.27 & 17.1 & 213 & 2.64 \\
     \hline
     binary fit & -- & -- & -- & 0.325 & 18 & 207.5 & 19.19 \\
     \hline
     triple system fit & $0.53\pm0.24$ & $3.13\pm0.21$ & $4.49\pm3.36$ & $0.25\pm0.04$ & $17.03\pm0.30$ & $210.37\pm1.72$ & 3.69 \\
     \hline
    \end{tabular}
   \end{center}
\end{table}

\subsubsection{Independant model fitting}

\begin{figure}[tp]
  \begin{center}
    \begin{tabular}{cc}
      \includegraphics[width=0.45\textwidth]{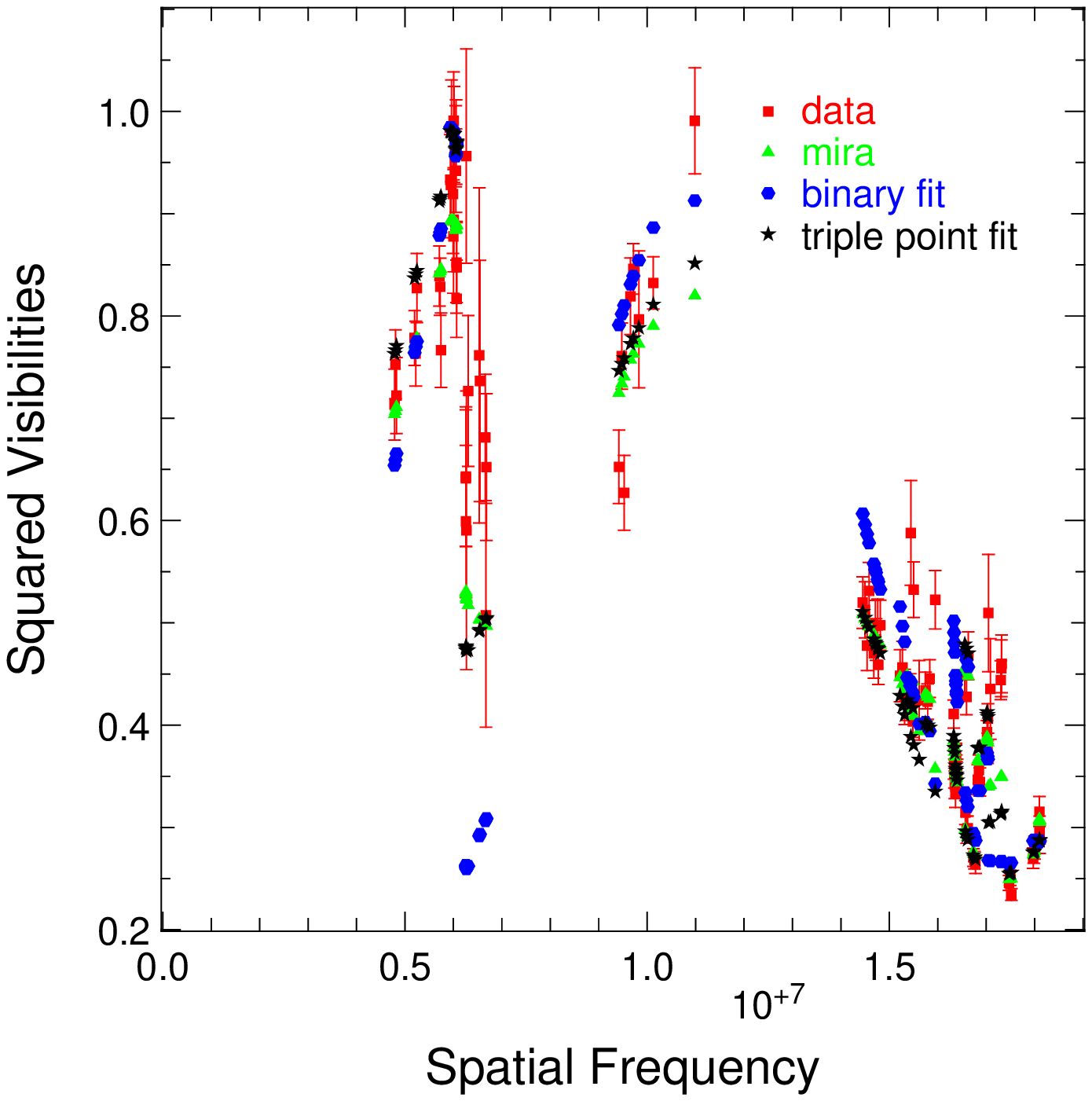} &
      \includegraphics[width=0.45\textwidth]{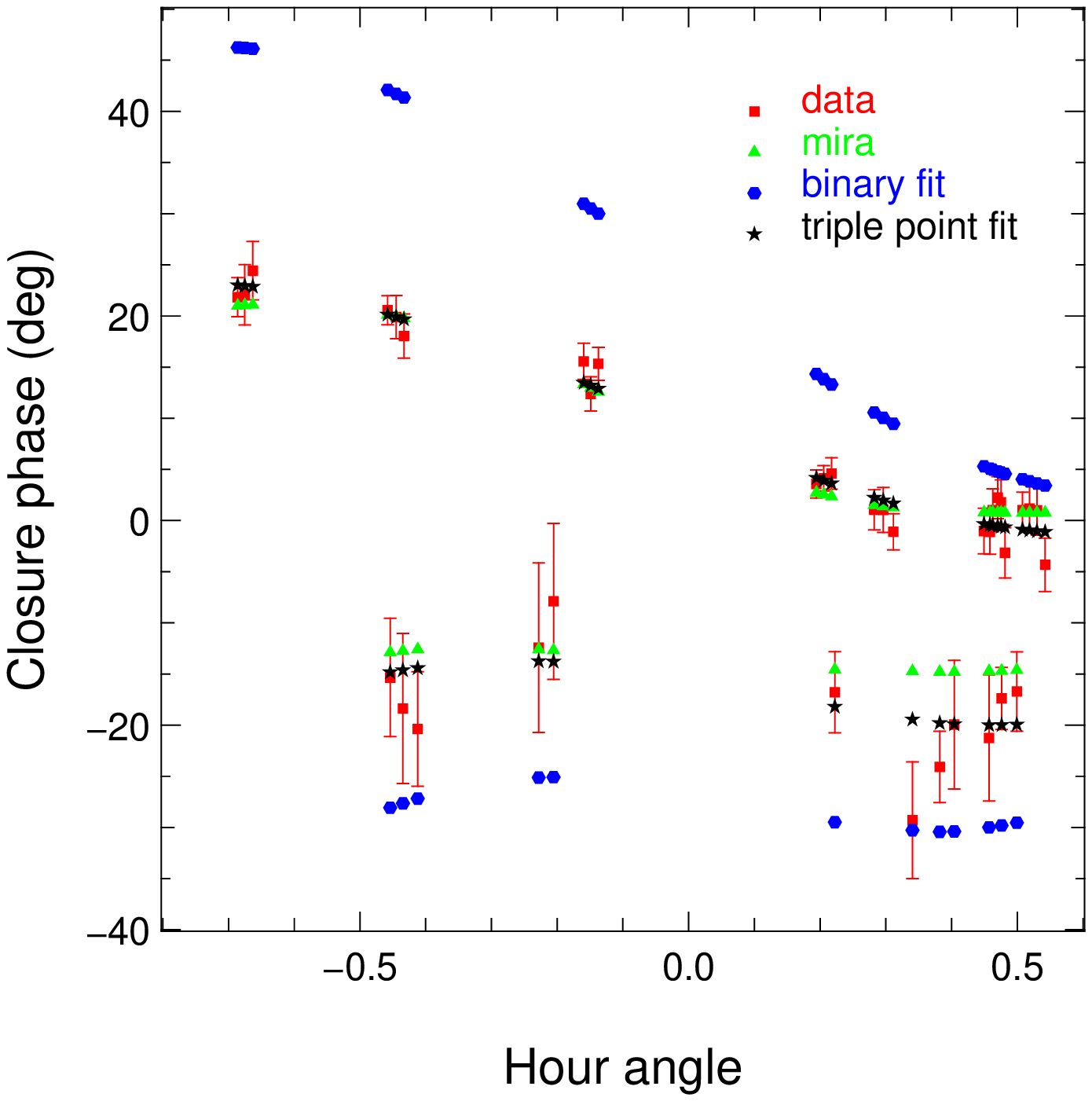}
    \end{tabular}
  \end{center}
  \caption{\label{fig:gwori_graph} Graphical result of the different methods used on GW Orionis. Left : the squared visibilities; Right : the closure phases. In red squares, the interferometric data; in green triangles, the reconstructed interferometric observables from the MIRA reconstructed image; in blue circles, the binary fit; in black stars, the triple point fit.}
\end{figure}

\begin{figure}[tp]
  \begin{center}
    \begin{tabular}{c}
      \includegraphics[width=0.5\textwidth]{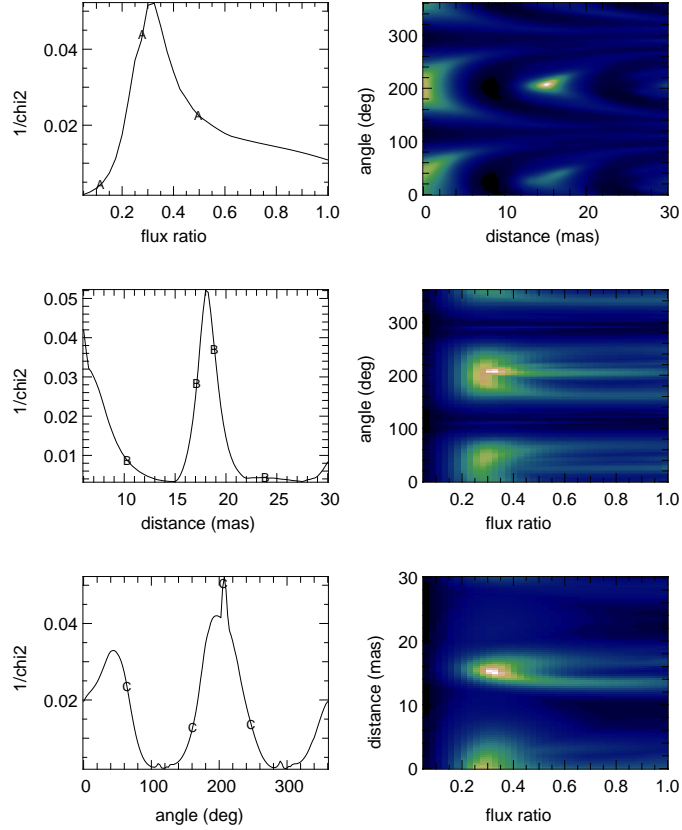} 
    \end{tabular}
  \end{center}
  \caption{\label{fig:gwori_chi2} Graphical representations of the
    $\chi^2$ cube between the GW Ori data and a binary model for a
    whole range of values for the three parameters : the flux ratio,
    distance and position angle.}
\end{figure}

In order to evaluate our confidence on the image reconstruction, we
decided to perform a model fitting totally independant of MIRA results
and to convince ourselves that the both methods lead to the same
results. We followed a two-step fitting procedure :

\begin{enumerate}
\item Binary fit : in a first step we carry on a grid search of the
  minimum $\chi^2$. For that purpose we compute an hypercube of
  minimum $\chi^2$ exploring all realistic binary parameters. From the
  resulting $\chi^2$ hypersurface, we determine an approximate
  position for the global minimum (see Table~\ref{tab:gwori_result},
  line 2 and Fig.~\ref{fig:gwori_graph} in blue circles). The
  graphical representations of the $\chi^2$ cube (see
  Fig.~\ref{fig:gwori_chi2}) allow us to evaluate the sharpness of the
  1/$\chi^2$ peak and the possible occurence of local minima; the curvature of the chi-2 around the minimum also provide estimation of the precision of the parameters.  As
  shown Table~\ref{tab:gwori_result} (line 2), the $\chi^2$ of this
  method is not satisfying.
\item Triple point fit : in order to improve the $\chi^2$ and
  motivated by the spectroscopic clear companion evidence, we add a
  third component to the model. This second step consists in a
  classical Levenberg-Marquardt minimisation of the $\chi^2$ with the
  six free parameters, three for each binary considering one component
  as the reference. The result of the binary fit is used as initial
  guess for the wide binary's parameters, whereas the parameters of
  the tight binary are choosen randomly. The LM fit result is shown
  Table~\ref{tab:gwori_result} (line 3) and graphically
  Fig.~\ref{fig:gwori_graph} (in black stars).
\end{enumerate}

We see that the result for the wide binary is approximately the same with both methods, whereas the results for the tight binary are quite different. The difference can be explained by the limited resolution of the interferometer and also by the fact that our data constrain  most the long separation than the shorter one due to the longer of baselines used for the observations.

\subsection{HD 45677} 
\label{sec:hd45677}

\begin{table}[t]
 \caption{Characteristics of the B$_{e}$ star HD 45677.}
  \label{tab:hd45677_caract}
   \begin{center}
    \begin{tabular}{ccccccccc}
    \hline
    \hline
     RA (J2000) & Dec (J2000) & m$_V$ & m$_J$ & m$_H$ & m$_K$ & Dist. (pc) & Spec. type \\
     \hline
     6h28m17.4s & -13\degre3'11.1'' & 8.05 & 7.2 & 6.3 & 4.8 & 350 & Bpshe \\
     \hline
    \end{tabular}
   \end{center}
\end{table}

\begin{figure}[tp]
  \begin{center}
    \begin{tabular}{cc}
      \includegraphics[width=0.45\textwidth]{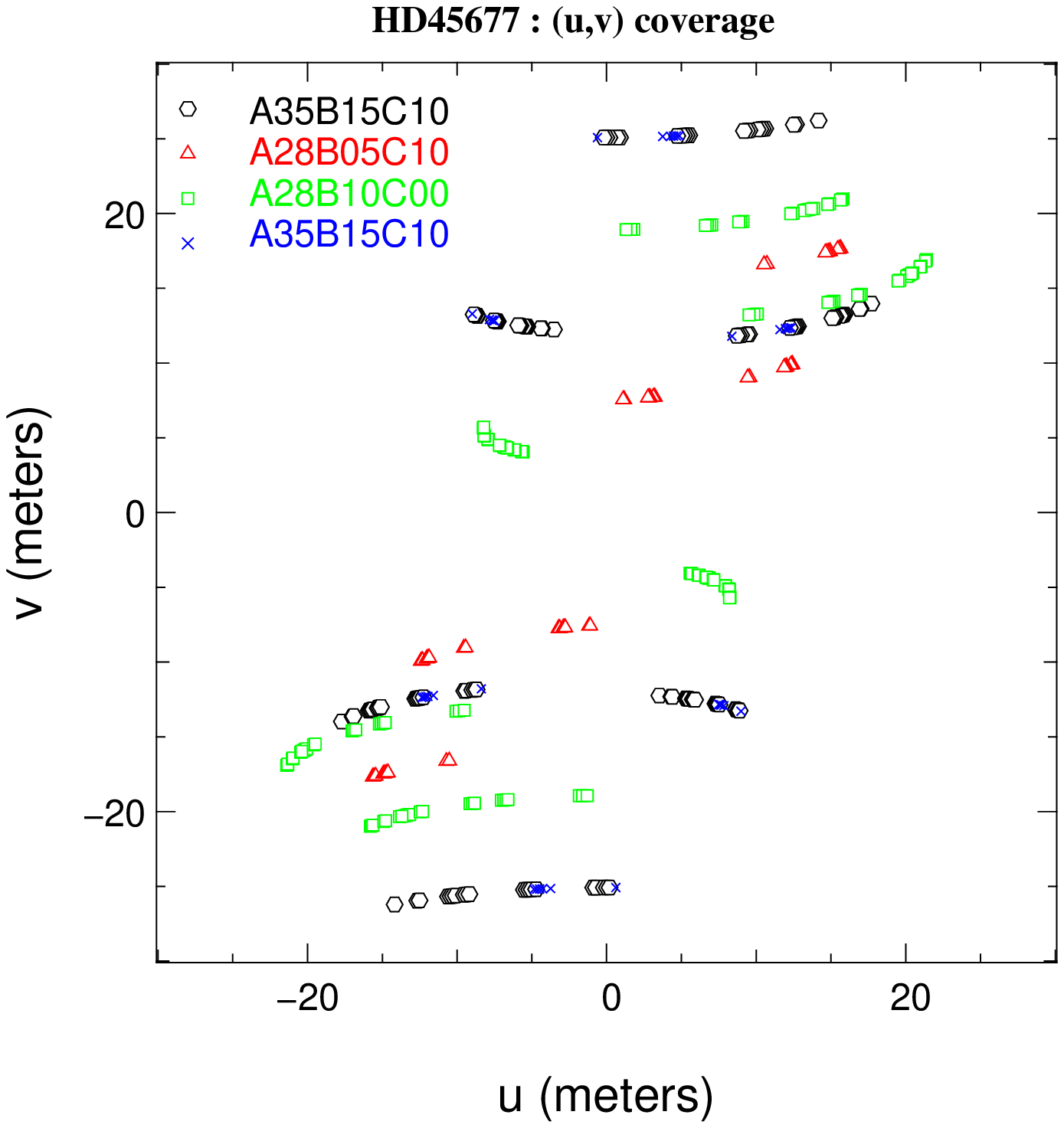} &
      \includegraphics[width=0.5\textwidth]{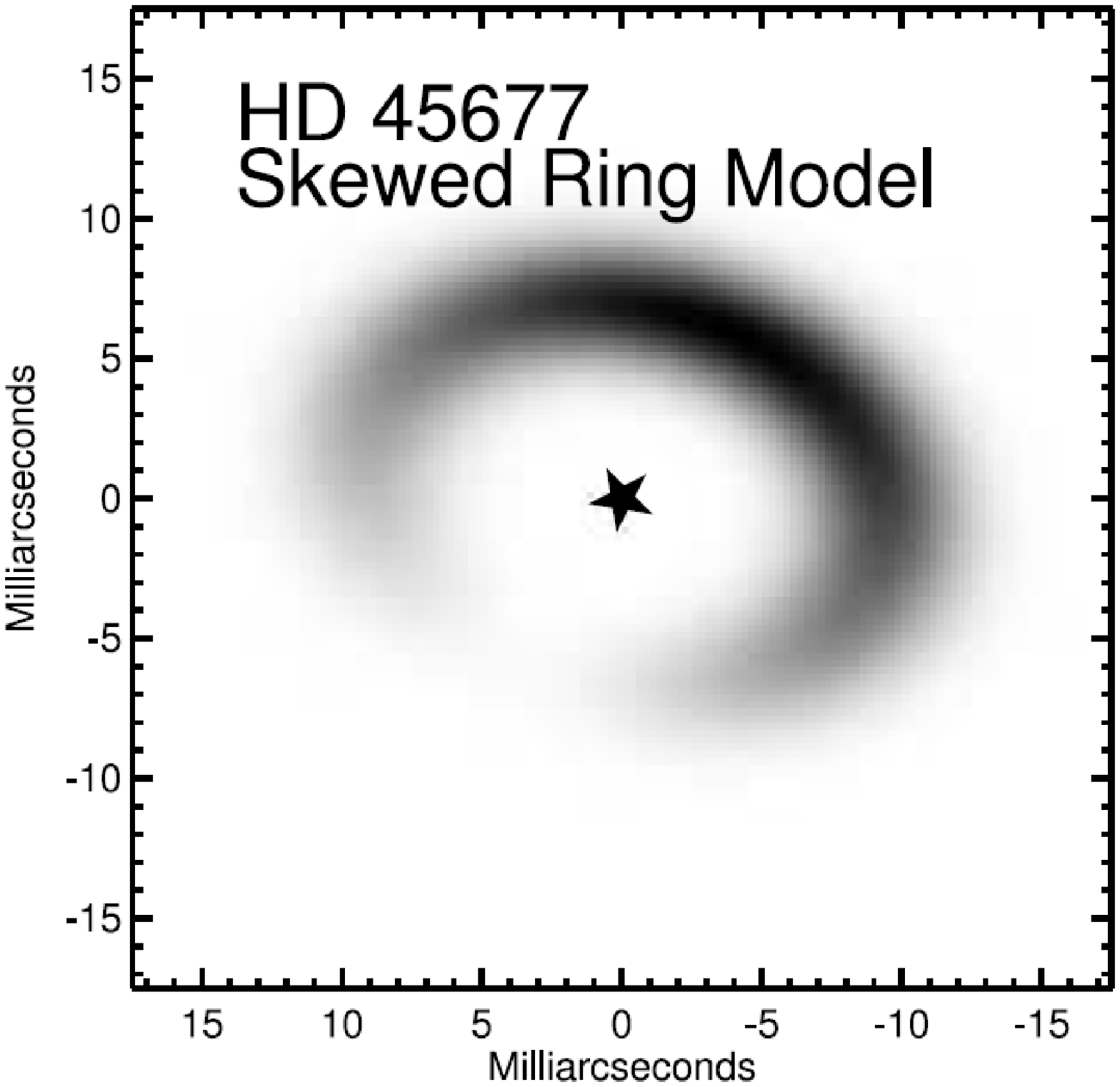} 
    \end{tabular}
  \end{center}
  \caption{\label{fig:hd45677_data}  
     $(u,v)$ coverage of the data set
    and the parametric imaging of IOTA data for HD 45677\cite{Monnier06}.}
\end{figure}

HD 45677 is a YSO that belongs to the Ae/Be Herbig group (see its
characteristics Table~\ref{tab:hd45677_caract}).  The first data of HD
45677 have been published by Monnier et al\cite{Monnier06}, given the
first closure phase survey of YSOs using an infrared interferometer.
In this paper, they did parametric imaging on HD 45677, proposing this
B$_{e}$ star to be surrounded by an elongated and highly skewed dust
ring (see then right panel of Fig.~\ref{fig:hd45677_data}).

\begin{figure}[tp]
  \begin{center}
    \begin{tabular}{c}
      \includegraphics[width=0.7\textwidth]{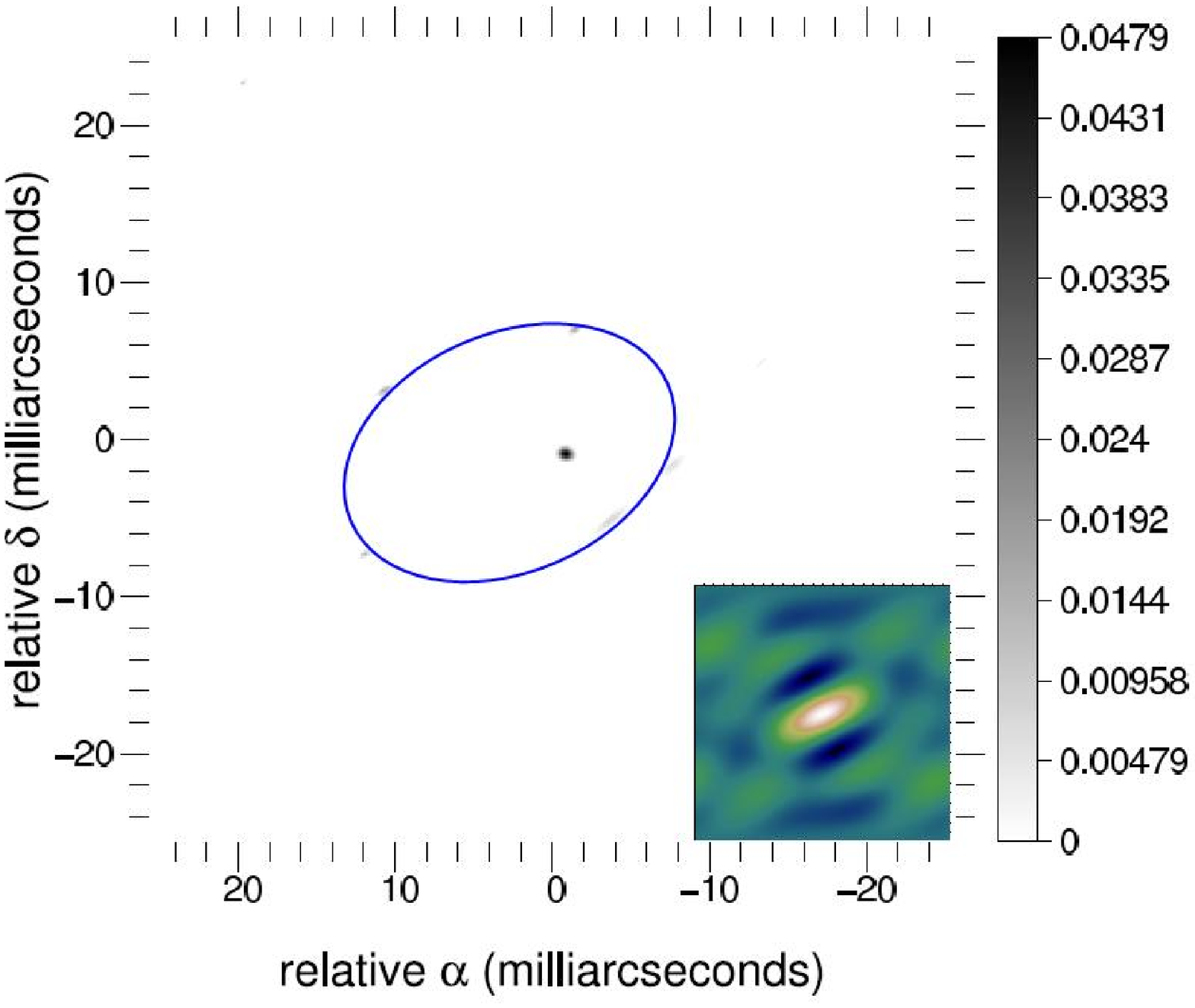}
    \end{tabular}
    \begin{tabular}{cc}
      \includegraphics[width=0.4\textwidth]{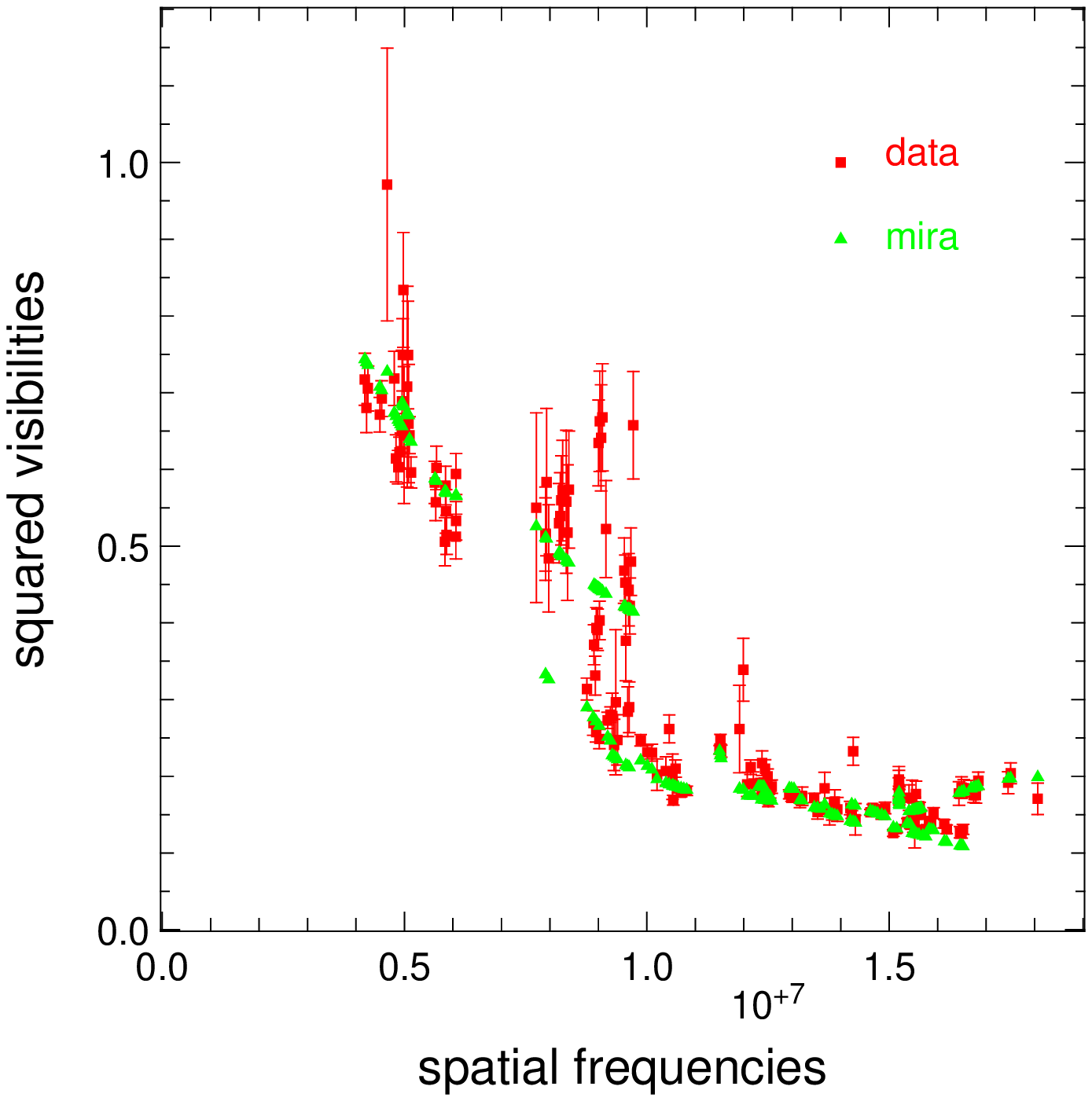} &
      \includegraphics[width=0.4\textwidth]{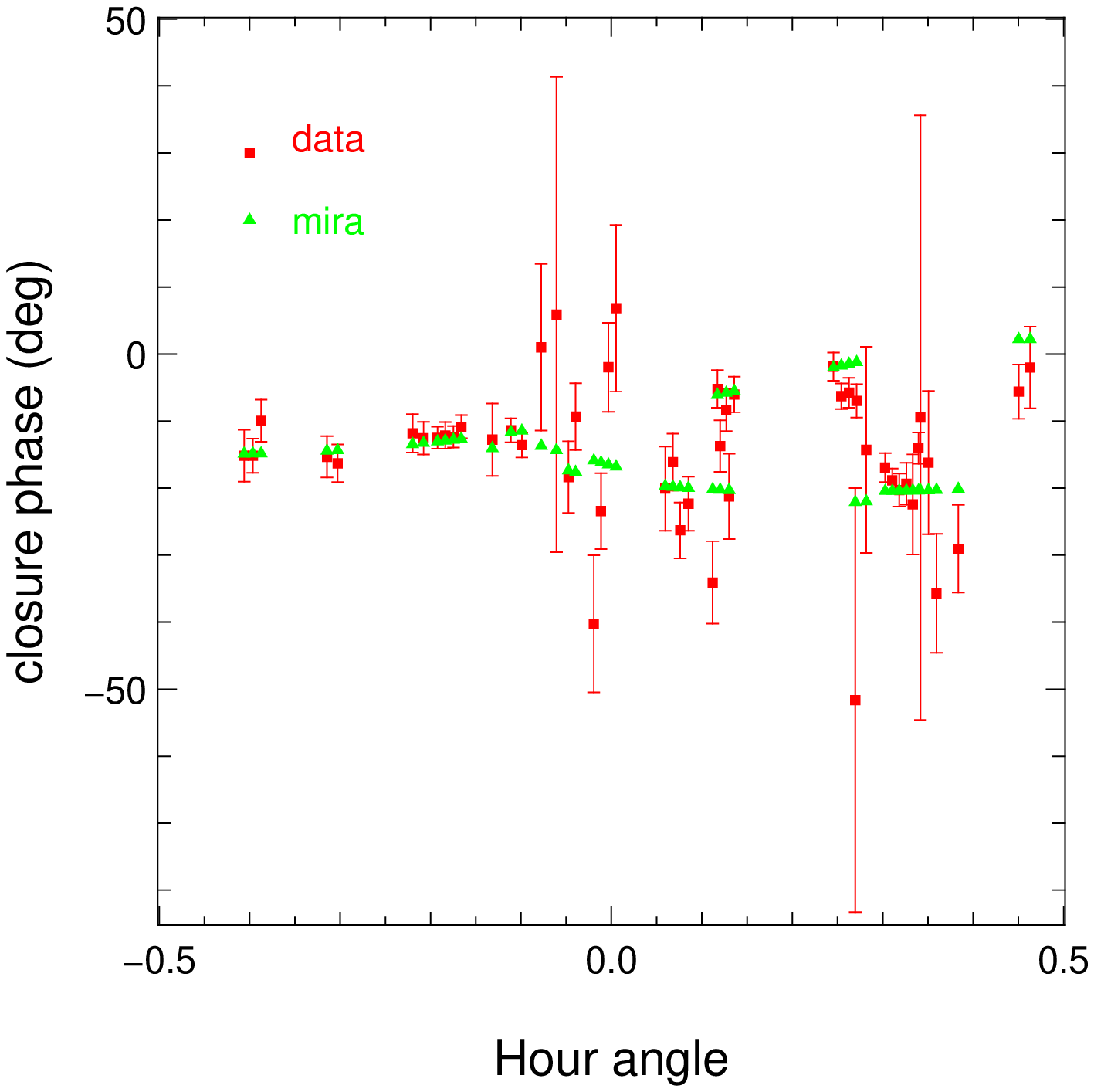} 
    \end{tabular}
  \end{center}
  \caption{\label{fig:hd45677_results} The result of the reconstructed image by MIRA for HD 45677 (with the dirty map). Second line left : in red squares, the measured squared visibilities; in green triangles, the squared visibilities of the reconstructed image. Second line right : idem for the closure phases.}
\end{figure}

The interferometric data of HD 45677 used for the reconstruction are
those of Monnier et al.\cite{Monnier06} (see Fig.~\ref{fig:hd45677_results} in red for the data and Fig.~\ref{fig:hd45677_data} for the $(u,v)$ plane) which have been obtained also
on the IOTA interferometer, as GW Orionis (see Sect.~\ref{sec:gwori})

The image reconstruction of HD 45677 has been done with the parameters below : 
\begin{itemize}
 \item a pixel size of 0.2 mas
 \item an image width of 256 pixels
 \item a regularization of type ``smoothness''
\end{itemize}

The result are showed Fig.~\ref{fig:hd45677_results}. The image seems
to show some points of a skewed disk, agreeing with the results obtain
by Monnier et al\cite{Monnier06}. To obtain a more precise image, we
need more interferometric measurements, especially at longer
baselines. This is why we requested several nights of observations at
VLTI with different triplets of baselines in order to have a good
coverage of the $(u,v)$ plan.

\section{CONCLUSION AND FUTURE}
\label{sec:conclu}

With only a few algorithms in the world, the image reconstruction in optical interferometry is really at its first steps especially for YSOs. The problem is much more complex than the image reconstruction in radio interferometry: in fact, in optical interferometry, we do not measure the complex visibility directly, the presence of the atmosphere leads to a lack of information on the phase and finally we have less measurements at the same time. Therefore, the optical interferometry needs its own method to retrieve images from the interferometric observables (squared visibilities and closure phases).

With one of these algorithms, MIRA, we show in this paper the first images from real long-baselines interferometric data of the environment of young stellar objects. The results have been analyzed with a new approach: we foung good correspondence between the reconstructed image and the parametric images, especially for GW Orionis. This demonstrates that this technique allows us to raise the degenerancy which appears when we analyse the interferometric data with model fitting.
However, as the image reconstruction is too heavy to do statistics on the reconstructed image and as there is not a method yet which can compute a significant error in order to estimate the rightness of the reconstructed image, we have difficulties to easily derive scientifical results from the reconstructed image.

\begin{figure}[tp]
  \begin{center}
\begin{tabular}{c}
 \includegraphics[width=0.4\textwidth]{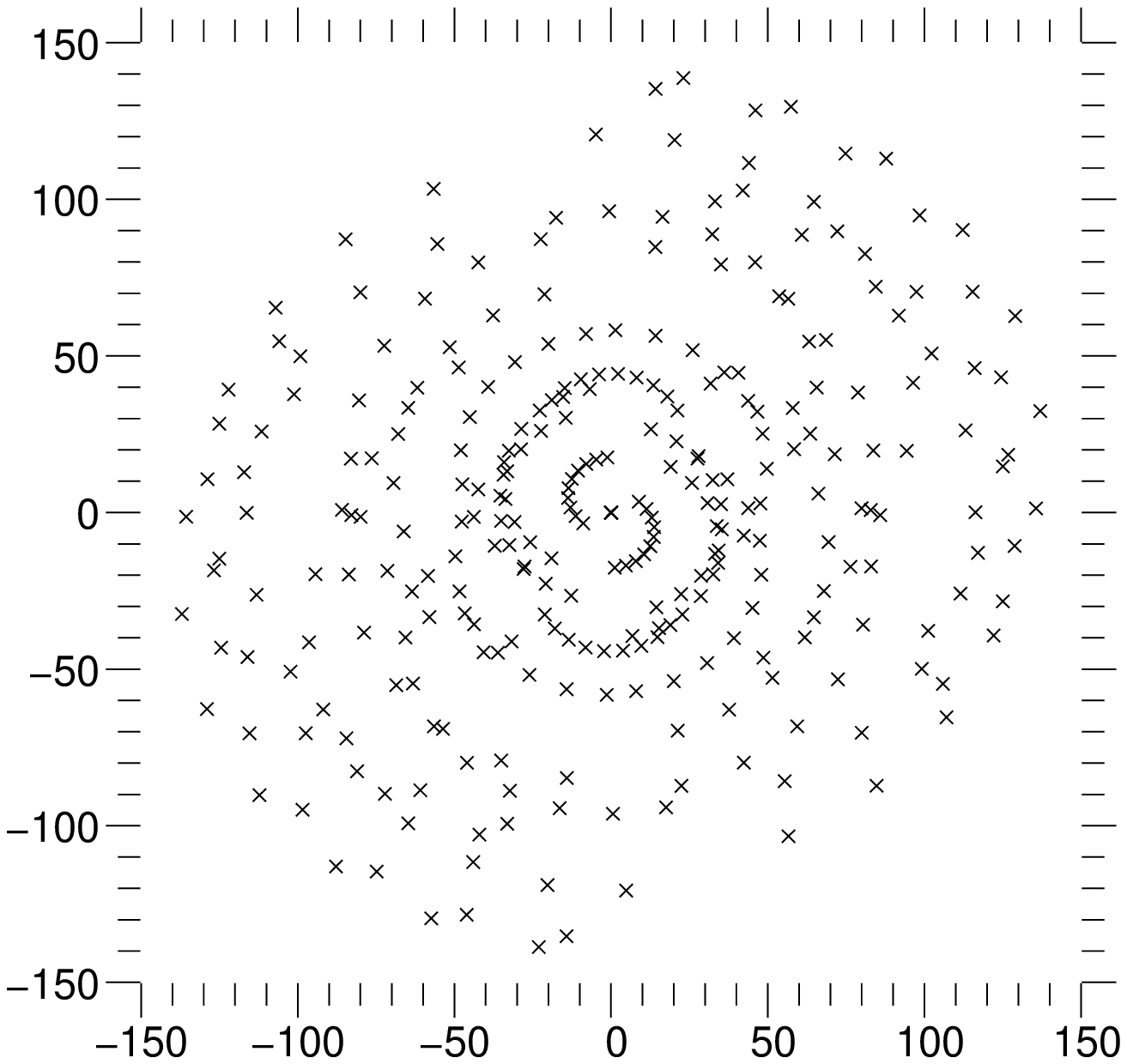}
\end{tabular}
    \begin{tabular}{cc}
      \includegraphics[width=0.4\textwidth]{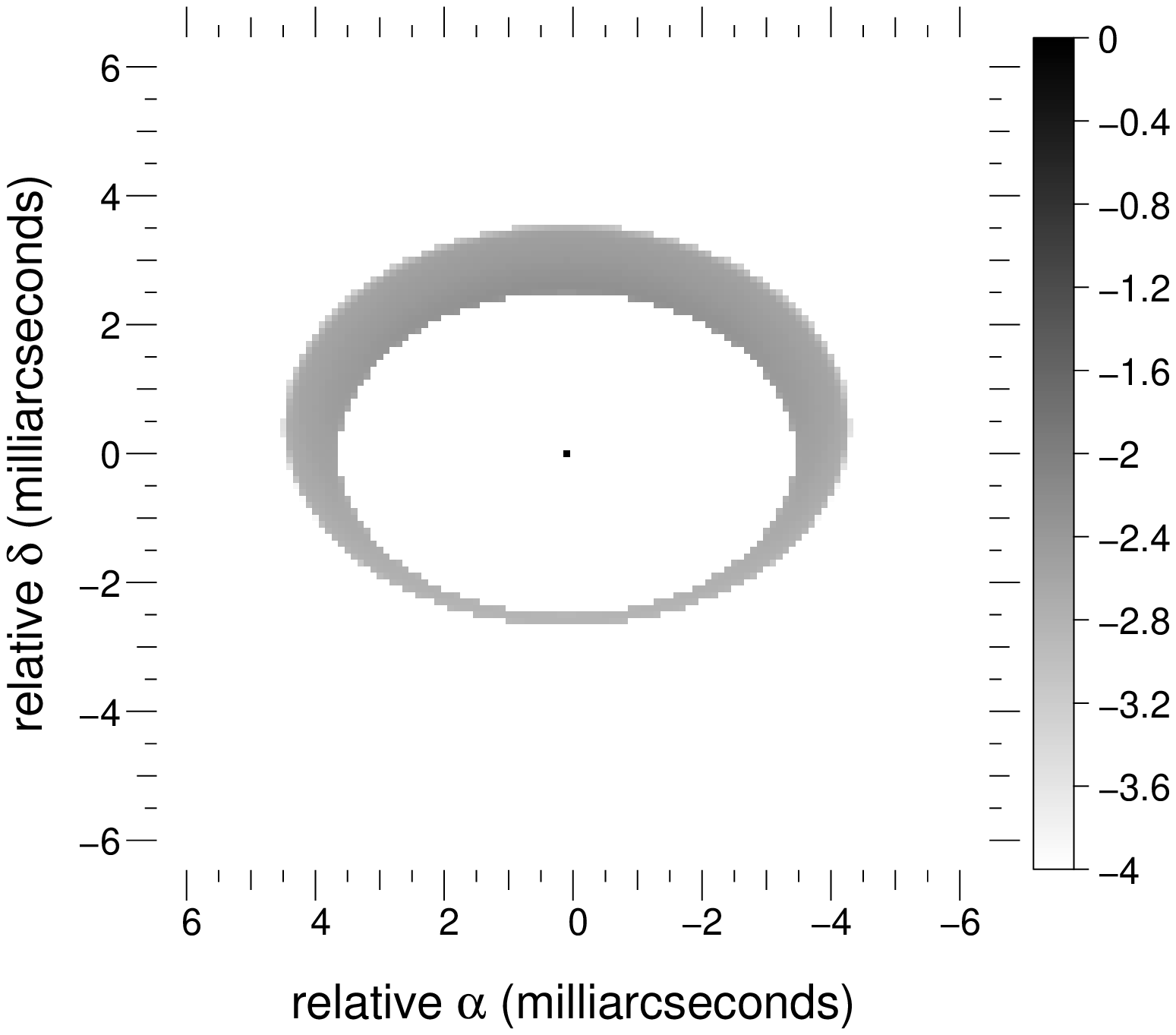} &
      \includegraphics[width=0.41\textwidth]{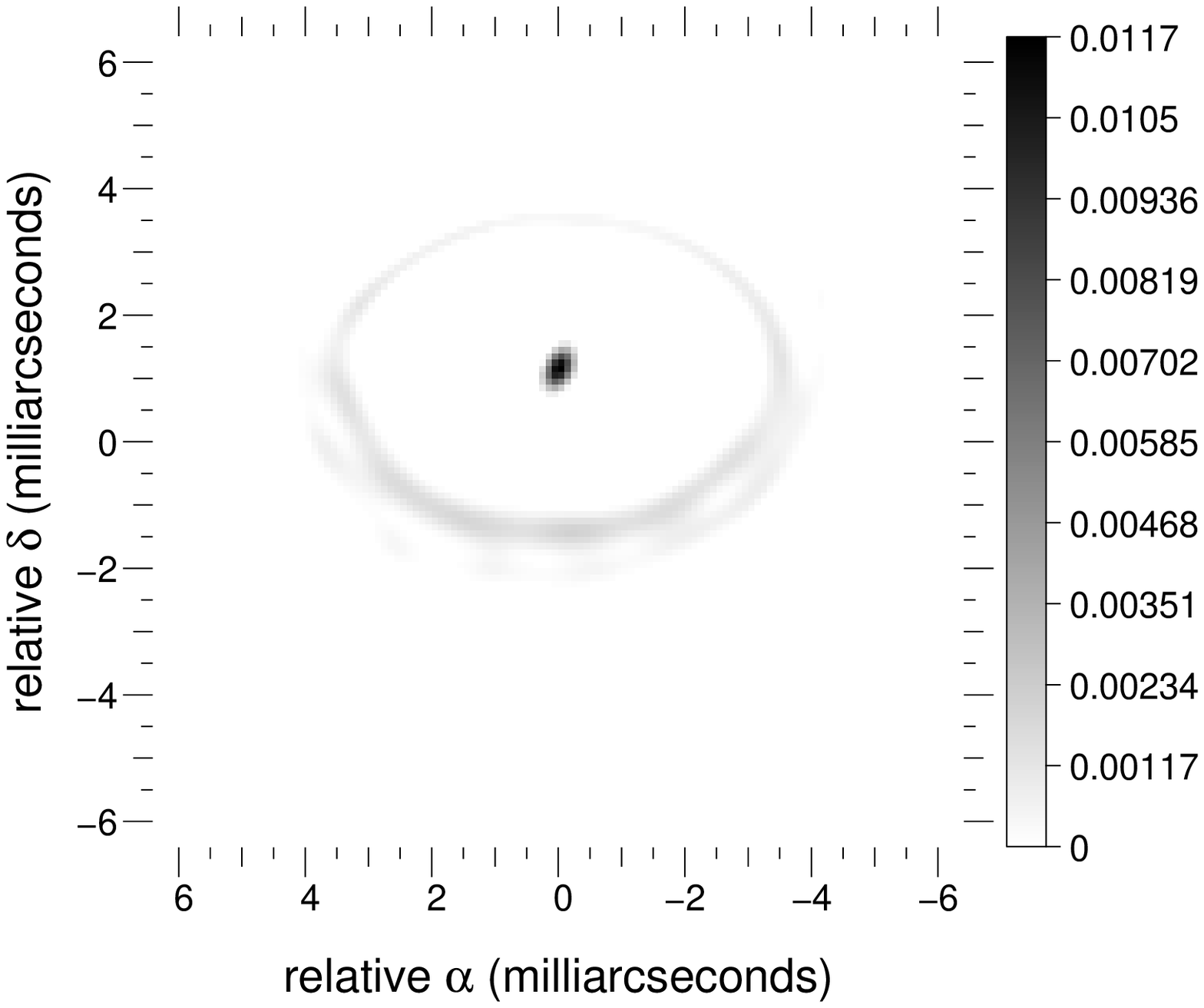} 
    \end{tabular}
\begin{tabular}{cc}
      \includegraphics[width=0.4\textwidth]{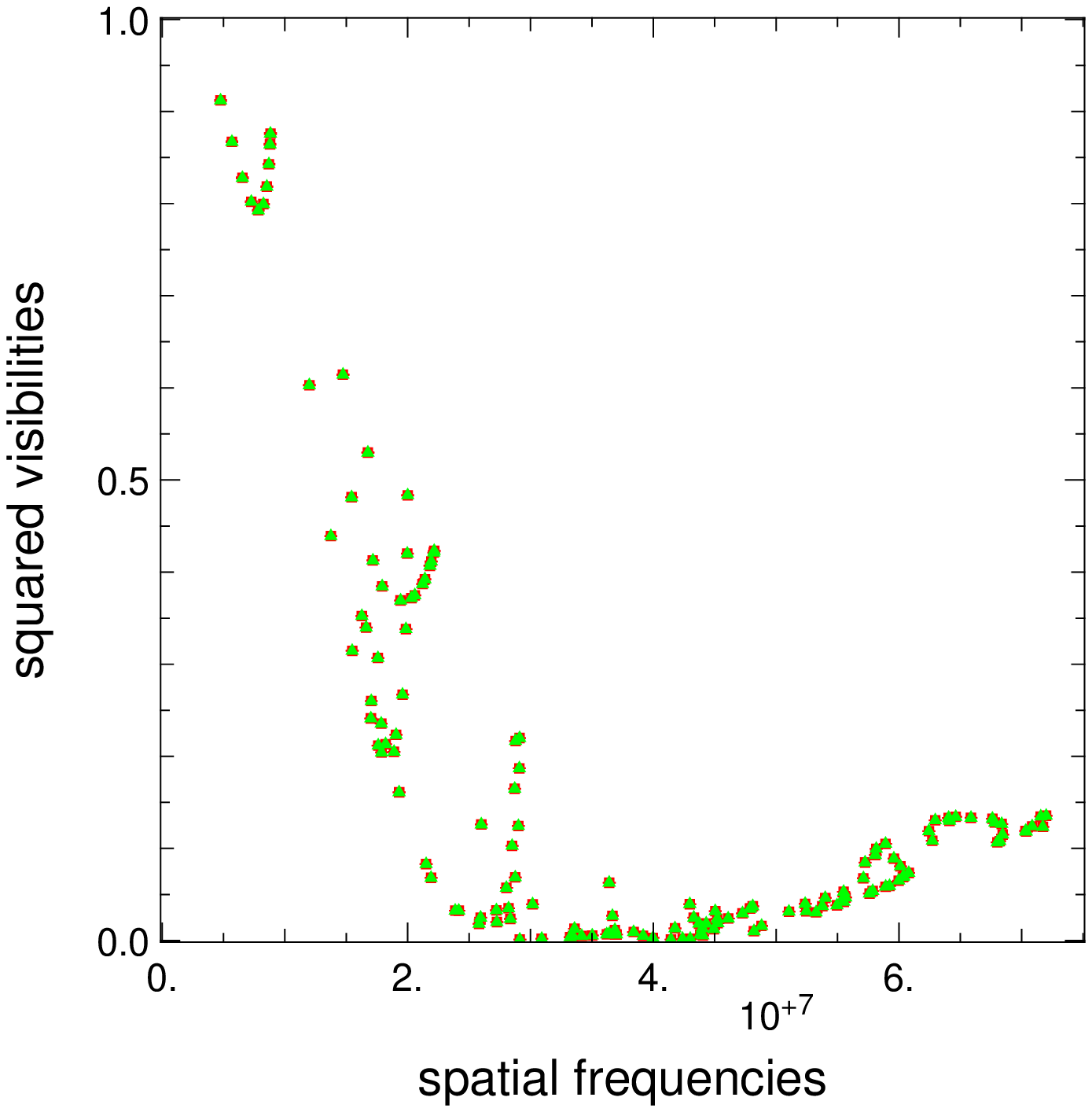} &
      \includegraphics[width=0.4\textwidth]{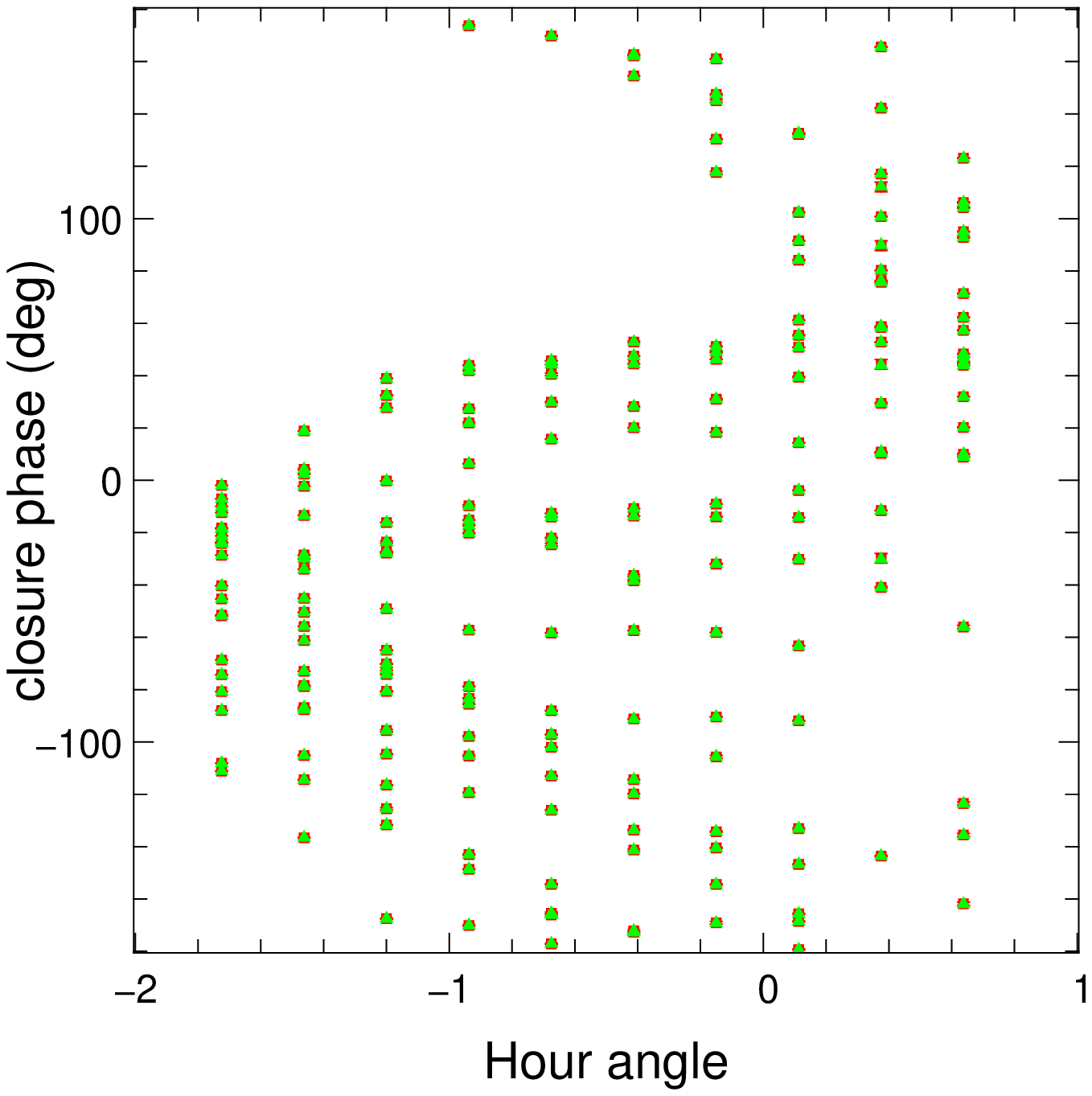} 
    \end{tabular}
  \end{center}
  \caption{\label{fig:VSI} Reconstructed image from simulated data for the science case of the VLTI Spectro-Imager (VSI): YSOs. First line: the $(u,v)$ coverage of the data set. Second line: the simulated image (left, logarithmic scale) and the reconstructed image by MIRA (right). Third line: the squared visibilities (left) and the closure phases (right) in red squared for the data and in green triangle for the MIRA results.}
\end{figure}

The image reconstruction in optical interferometry needs therefore to make progress but it already shows that the technique will give great facilities in the analysis of the interferometric data. With future instrument as the VLTI Spectro Imager VSI, a second generation instrument for the VLTI, we can make image reconstruction of several astronomical objects\cite{FilhoSPIE}  : with the possibility of observing with 4 to 6 telescopes and with a groovy decrease of the error bars, the $(u,v)$ plane coverage is well much better and the reconstruction made by MIRA is easier (see Fig.~\ref{fig:VSI} for an example on YSOs)

\bibliography{SPIEtalk_ok.bib}   
\bibliographystyle{spiebib}   

\end{document}